\documentclass[review,numbers,sort,compress]{elsarticle}

\usepackage{lineno}
\usepackage{graphicx}
\usepackage{latexsym}%
\usepackage[table]{xcolor}
\usepackage{amsfonts}
\usepackage{booktabs}
\usepackage{lmodern}
\usepackage{soul}
\usepackage{enumitem}
\usepackage{comment}
\usepackage{relsize}
\usepackage{lipsum}
\usepackage{color}
\usepackage{siunitx}
\usepackage[breaklinks=true]{hyperref}
\usepackage{breakcites}
\usepackage{blindtext, rotating}
\usepackage{framed} 
\usepackage{multicol} 
\usepackage{pifont}
\usepackage[parfill]{parskip}
\usepackage{nomencl}
\usepackage{colortbl}
\usepackage{tabularx}
\usepackage[all]{nowidow}
\usepackage{amssymb}
\usepackage{amsmath}
\usepackage{epstopdf}
\usepackage{setspace}
\usepackage{xspace}
\usepackage{subcaption}
\usepackage{epsfig}
\usepackage{placeins}
\usepackage{epstopdf}
\usepackage{wrapfig}
\usepackage{algorithm}
\usepackage[noend]{algpseudocode}
\usepackage{caption}
\usepackage[version=3]{mhchem}
\usepackage{lineno,hyperref}
\usepackage{cleveref}
\usepackage[font=small,skip=0pt]{caption}
\usepackage{geometry}
\usepackage{todonotes}

\renewcommand{\linenumbers}{}

\modulolinenumbers[5]
\definecolor{lavender}{rgb}{0.75, 0.58, 0.89}

\newcolumntype{M}[1]{>{\centering\arraybackslash}p{#1}}
\newcolumntype{P}[1]{>{\raggedright\arraybackslash}p{#1}}
\DeclareRobustCommand{\vardbtilde}[1]{\tilde{\raisebox{0pt}[0.9\height]{$\tilde{#1}$}}}
\newcommand{\prt}{\partial}
\newcommand{\bs}{\mbox{\boldmath $s$}}
\newcommand{\baa}{\mbox{\boldmath $a$}}
\newcommand{\bbb}{\mbox{\boldmath $b$}}

\newcommand{\bchi}{\mbox{\boldmath $\chi$}}

\newcommand{\bv}{\mbox{\boldmath $v$}}
\newcommand{\bu}{\mbox{\boldmath $u$}}
\newcommand{\bw}{\mbox{\boldmath $w$}}
\newcommand{\by}{\mbox{\boldmath $y$}}
\newcommand{\bbf}{\mbox{\boldmath $f$}}
\newcommand{\bxi}{\mbox{\boldmath $\xi$}}
\newcommand{\bpsi}{\mbox{\boldmath $\psi$}}
\newcommand{\bP}{\mbox{\boldmath $p$}}
\newcommand{\balpha}{\mbox{\boldmath $\alpha$}}
\newcommand{\lp}{\left(}
\newcommand{\rp}{\right)}

\setlist[itemize]{leftmargin=*}

\newif\ifcomments 
\commentsfalse
\commentstrue 

\definecolor{Green}{rgb}{0,0.5,0}

\modulolinenumbers[5]

\definecolor{lightgray}{gray}{0.9}
\definecolor{amethyst}{rgb}{0.8, 0.0, 0.8}
\definecolor{aogreen}{rgb}{0.01, 0.75, 0.24}

\journal{Journal of \LaTeX\ Templates}

\begin{document}

\begin{frontmatter}

\title{Skeletal Model Reduction with Forced Optimally Time Dependent Modes}
\tnotetext[mytitlenote]{Fully documented templates are available in the elsarticle package on \href{http://www.ctan.org/tex-archive/macros/latex/contrib/elsarticle}{CTAN}.}

\author[mymainaddress]{A.G.\ Nouri\corref{mycorrespondingauthor}}
\author[mymainaddress]{H.\ Babaee}
\author[mymainaddress]{P.\ Givi}
\author[mysecondaryaddress]{H.K.\ Chelliah}
\author[myteritaryaddress]{D.\ Livescu}

\address[mymainaddress]{Department of Mechanical Engineering and Materials Science, University of Pittsburgh, Pittsburgh, PA 15261, USA}
\address[mysecondaryaddress]{Department of Mechanical and Aerospace Engineering, University of Virginia, Charlottesville, VA 22904, USA}
\address[myteritaryaddress]{Los Alamos National Laboratory, Los Alamos, NM 87544, USA}

\cortext[mycorrespondingauthor]{Corresponding author}

\begin{abstract}
Sensitivity analysis with forced  optimally time dependent (f-OTD) modes is introduced and its application for skeletal model reduction is demonstrated. f-OTD expands the sensitivity coefficient matrix into a low-dimensional, time dependent, orthonormal basis which captures directions of the phase space associated with most dominant sensitivities. These directions highlight the instantaneous active species and reaction paths. Evolution equations for the orthonormal basis and the projections of sensitivity matrix onto the basis are derived, and the application of f-OTD for skeletal reduction is described. In this framework, the sensitivity matrix is modeled, stored in a factorized manner, and never reconstructed at any time during the calculations. For demonstration purposes, sensitivity analysis is conducted of constant pressure ethylene-air burning in a zero-dimensional reactor and new skeletal models are generated. The flame speed, the ignition delay, and the extinction curve of resulted models are compared against some of the existing  skeletal models. The results  demonstrate the ability of f-OTD approach to eliminate unimportant reactions and species in a systematic, efficient and accurate manner.
\end{abstract}


\end{frontmatter}

\linenumbers

\section{Introduction}
\label{section:Intro}
Detailed reaction models for C$_1$-C$_4$ hydrocarbons usually contain over $100$ species in about $1000$ elementary reactions~\cite{EC11,SYWL09,JetSurf,UCSD_EthylAir,CRECK,AramcoMech2}. Direct application of such models is only limited to simple, canonical combustion simulations because of their tremendous computational cost. Therefore, various reduction techniques have been developed to accommodate realistic fuel chemistry simulations, and to capture intricacies of chemical kinetics in complex multi-dimensional combustion simulations. As the first step in developing model reduction, it is important to extract a subset of the detailed reaction model, \textit{skeletal model}, by eliminating unimportant species and  reactions in a systematic manner~\cite{Smooke91,Peters1993}. For example, local sensitivity analysis (SA) and reaction flux analysis are often used in  reduction methods \cite{Turanyi90a,wang:1991,SCGJ10}. Other approaches such as directed relation graph (DRG) and its variants have also been used~\cite{LL05,ZLL07,LL09,NSR10}. Global sensitivity analyses are useful for studying   uncertainty of kinetic parameters (\textit{i.e.} collision frequencies and activation energies) propagate through model and non-linear coupling effects~\cite{Turanyi90b,Saltelli04,SYWL09,ESC12}. Local SA, which is the subject of this work, explores the response of the model output to a small change of a parameter from its nominal value~\cite{Shuang16}, and global SA estimates the effect of input parameters across the whole uncertainty space of model parameters on predictions of interest~\cite{Sobol90,HS96,RA99,Sobol01,LSR02}. 

Model reduction with local SA contains methods such as principal component analysis (PCA) \cite{VVT85,brown:1997} and construction of a species ranking~\cite{SFCFR16}. In local SA, the sensitivities are commonly computed via finite difference (FD) discretization, or by directly solving a sensitivity equation (SE), or by an adjoint equation (AE)~\cite{DCB20}. The computational cost of using FD or SE, which are \textit{forward} in time methods, scales linearly with the number of parameters – making them impracticable when sensitivities with respect to a large number of parameters are needed. On the other hand, computing sensitivities with AE requires a \textit{forward-backward} workflow, but the computational cost is independent of the number of parameters as it requires solving a single ordinary/partial differential equation (ODE/PDE)~\cite{BOR15,LCRPS19,Langer2020}. The AE solution is tied to the objective function, and for cases where multiple objective functions are of interest, the same number of AEs must be solved for each objective function. Regardless of the method of computing  sensitivities, the output of FD, SE, and AE at each time instance is the full sensitivity matrix, which can be extremely large for systems with very large number of parameters.

\begin{figure}[!htbp]
\centering
 \epsfig{file=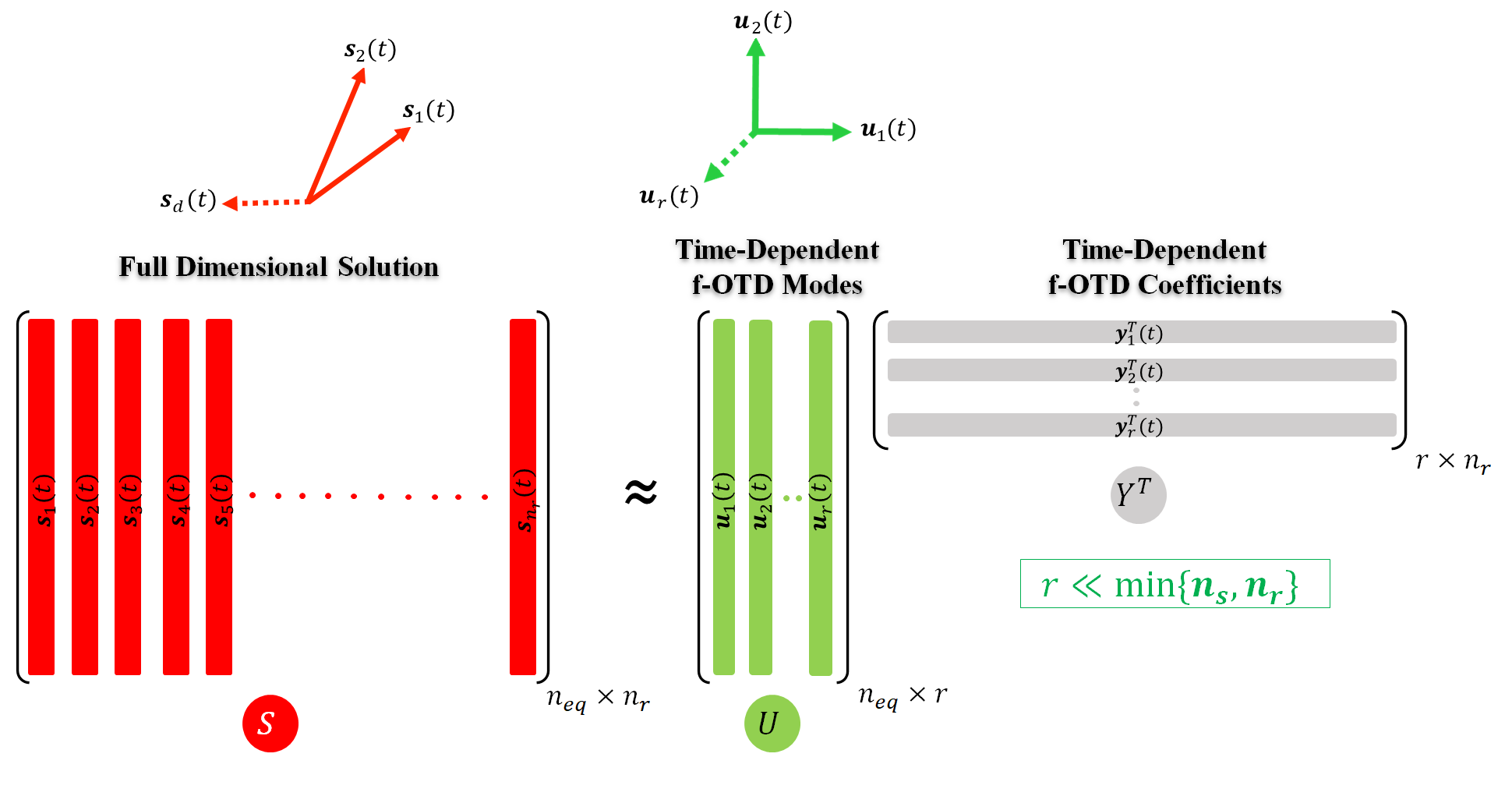,width=14.0cm}
 \caption{Modeling sensitivity matrix $S(t)$ as a multiplication of two low-ranked matrices $U(t)$ and $Y(t)$ which evolve based on Eqs.\ (\ref{eq:U_evolution}) and (\ref{eq:Y_evolution}).}
\label{FIG:S_matrix_shape}
\end{figure}

Recently, forced optimally time dependent (f-OTD) decomposition was introduced for computing sensitivities in evolutionary systems using a \textit{model driven} low-rank approximation~\cite{DCB20}. This methodology is the extension of OTD decomposition in which a mathematical framework is laid out for the extraction of the low-rank subspace associated with transient instability of the dynamical system~\cite{BS16}. In \textit{forward} workflow of f-OTD, the sensitivity matrix \textit{i.e.} $S \in \mathbb{R}^{n_{eq} \times n_r}$ is modeled on-the-fly as the multiplication of two skinny matrices $U(t) = [\bu_1(t), \bu_2(t), \dots, \bu_r(t)] \in \mathbb{R}^{n_{eq} \times r}$, and $Y(t) = [\by_1(t), \by_2(t), \dots, \by_r(t)] \in \mathbb{R}^{n_r \times r}$ which contain the f-OTD modes and f-OTD coefficients, respectively (Fig.\ \ref{FIG:S_matrix_shape}), where $n_{eq}$ is the number of equations (or outputs), $n_r$ is the number of independent parameters, and $r \ll$ min\{$n_s, n_r$\} is the reduction size. The key characteristic of f-OTD is that both $U(t)$ and $Y(t)$ are time-dependent and they evolve based on closed form evolution equations extracted from the model, and are able to capture sudden transitions associated with the largest finite time Lyapunov exponents~\cite{BFHS17}. The time-dependent bases have also been used for the purpose of stochastic reduced order modeling \cite{SL09,CHZI13,Babaee:2017aa,B19,PB20} and recently for on the fly reduced order modeling of reactive species transport equation \cite{RNB21}. In a nutshell, f-OTD workflow i) is \textit{forward} in time unlike AE, ii) bypasses the computational cost of solving FD and SE, or other data-driven reduction techniques, and iii) stores the modeled sensitivities in a reduced format unlike FD, SE and AE. 

The major advantage of PCA in skeletal reduction is to combine the sensitivity coefficients for a wide range of operating conditions (\textit{e.g.} equivalence ratio and pressure) \cite{EC11}. 
Principal component analysis finds the low-dimensional subspace of data gathered from different (temporal or spatial) locations by applying a minimization algorithm over the whole data at once. Therefore, PCA is a low-rank approximation in a time-averaged sense and may fail to capture highly transient finite-time events   (\textit{e.g.} ignition) unless it is given enough data from locations associated with such phenomena. Pre-recognizing these important locations and selecting optimized number of locations needs knowledge and expertise. Moreover, as PCA modes are time invariant, the process of selecting  sufficient eigenvalues/eigenmodes to consider in PCA is crucial and usually done by trial and error. Principal component analysis usually finds one eigenmode (group of reactions) with the eigenvalue which is several orders of magnitude larger than the others. References~\cite{ZT03,EC11} shw that for certain problems, a skeletal model built solely upon the information conveyed by that first reaction group from PCA can fail to accurately reproduce the detailed model over the entire domain of interest. Therefore, one needs to deal with several eigenmodes with close eigenvalues and choose  essential reaction groups among them \cite{EC11}. 

In order to resolve the drawbacks of current SA methods and PCA for skeletal model reduction, we i) use f-OTD methodology for SA, and ii) present a new framework for observing the sensitivities in a \textit{compressed} format for skeletal reduction. The applicability of our approach for skeletal  reduction is demonstrated for ethylene-air burning based on the USC model~\cite{SYWL09}. Adiabatic, constant pressure, spatially homogeneous ignition is the canonical problem; and the generated skeletal models with f-OTD are benchmarked against detailed and skeletal models in the literature based on their predicted ignition delay, flame speed, and extinction curve.

The remainder of this paper is organized as follows. The theoretical description of PCA and f-OTD and their mathematical derivations for SA is presented in \S~2. Model reduction with f-OTD is first described in \S~3 with a simple reaction model for hydrogen-oxygen combustion, followed by skeletal model reduction with f-OTD for more complex ethylene-air system using the USC model in \S~4. The paper ends with conclusions in \S~\ref{section:conclu}. All the generated models  are furnished  in supplementary materials section.

\section{Formulation}\label{section:PCA_OTD}

Consider a chemical system of $n_s$ species reacting through $n_r$  irreversible reactions,
\begin{equation}
\sum_{k=1}^{n_s} \nu'_{kj} \mathbb{M}_{k} \rightarrow
\sum_{k=1}^{n_s} \nu''_{kj} \mathbb{M}_{k}, \ \ j=1, \dots n_r,
\end{equation}

\noindent where $\mathbb{M}_{k}$ is a symbol for species $k$, and $\nu'_{kj}$ and $\nu''_{kj}$ are the molar stoichiometric coefficients of species $k$ in reaction $j$. Changes of mass fractions $\bpsi = [\psi_1, \psi_2, \dots, \psi_{n_s}]^T$   and temperature $T$ in an adiabatic, constant pressure $p$, and spatially homogeneous reaction system of ideal gases can be described by the following initial value problems (IVPs) \cite{ZZT03}
\begin{subequations}
\label{eq:psi_T_IVP}
\begin{eqnarray}
 \frac{d\psi_k}{dt} &=& f_{\psi_k}(\bpsi,T,\balpha)  = \frac{W_k}{\rho} \sum_{j=1}^{n_r} \nu_{kj} \mathcal{Q}_j, \  \  \bpsi(0) = \bpsi_0,\\
 \frac{dT}{dt} &=& f_{T}(\bpsi,T,\balpha) = -\frac{1}{ c_p} \sum_{k=1}^{n_s}\Delta h^0_{f,k} f_{\psi_k},\ \ T(0) = T_0,  
\end{eqnarray}
\end{subequations}
\noindent where $t \in [0, t_{f}]$ is time, $t_f$ is the final time and $W_k$ and $\Delta h^0_{f,k}$ are the molecular weight and enthalpy of formation of species $k$, respectively, and 
\begin{subequations}
\label{eq:Q_nu}
\begin{eqnarray} 
 \nu_{kj} &=& \nu''_{kj} - \nu'_{kj}, \\
 \mathcal{Q}_j &=& \alpha_j k_{j} \prod_{k=1}^{N_s} \lp \frac{\rho\psi_k}{W_k} \rp^{\nu'_{kj}}.
\end{eqnarray}
\end{subequations}

\noindent Here, $\balpha = [1,1, \dots, 1] \in \mathbb{R}^{n_r}$ is the vector of parameters and $k_{j}$ is the rate constant which is usually modeled using the modified Arrhenius parameters~\cite{Williams1985} for elementary reactions (Note: all reversible reactions are cast as irreversible reactions). In Eq.\ (\ref{eq:psi_T_IVP}) $\rho(T,\bpsi)$ and $c_p(T,\bpsi)=\sum_{k=1}^{n_s} \psi_k c_{p_k}(T)$ are the density and specific heat at constant pressure of the mixture, respectively, where $c_{p_k}(T)$ is the specific heat at constant pressure of $k$th species  given by the NASA coefficient polynomial parameterization \cite{Cantera} and $\rho$ is given by the ideal gas equation of state. 
Let $\bxi = [\bpsi , T] \in \mathbb{R}^{n_{eq}}$ denote the vector of compositions and accordingly $\bbf = [\bbf_{\bpsi} , f_T]$ where $n_{eq} = n_s +1$. Then the compositions IVP would be,
\begin{equation}
\label{eq:baseODE}
\frac{d\xi_{i}}{dt} = f_{i} \lp \bxi,\balpha \rp, \ \ \bxi(0) = [\bpsi_0,T_0].
\end{equation}

\noindent Since $\balpha = \mathbf{1}$, the perturbation with respect to $\alpha_j$ amounts to an infinitesimal perturbation of progress rates $\mathcal{Q}_j$ for $j=1,2, \dots, n_r$.  The sensitivity matrix, $S(t)=[\bs_1(t), \bs_2(t), ... \bs_{n_r}(t)]\in \mathbb{R}^{n_{eq} \times n_r}$, contains local sensitivity coefficients, $\bs_{k}=\prt \bxi / \prt \alpha_k$, and it can  be calculated by solving the SE,
\begin{equation}
 \label{eq:sensit_exact}
 \begin{split}
\frac{dS_{ik}}{dt} &=   \frac{\prt f_{i}}{\prt \xi_{j}} \frac{\prt \xi_{j}}{\prt \alpha_{k}} + \frac{\prt f_{i}}{\prt \alpha_{k}}  = L_{ij} S_{jk} + F_{ik}, 
\end{split}
\end{equation}
\noindent where $L_{ij} =  \frac{\prt f_{i}}{\prt \xi_{j}}$ and $F_{ij}=\frac{\prt f_{i}}{\prt \alpha_{k}}$ are Jacobian and forcing matrices. 

\subsection{Principal component analysis}
\label{subsection:PCA_sec}

Principal component analysis investigates the effects of parameter perturbations on the objective function $\mathcal{G}(\bP)$

\begin{equation}
\label{eq:object_func}
\mathcal{G}(\bP) = \mathlarger{\int_{z_1}^{z_2}} \mathlarger{\sum}_{i=1}^{n_{eq}} \bigg[ \frac{\xi_i(z,\bP) - \xi_i(z,\bP^{\textbf{0}})}{\xi_i(z,\bP^{\textbf{0}})} \bigg]^2 dz,
\end{equation}

\noindent where $\bP^{\textbf{0}}$ and $\bP$ are unperturbed and perturbed normalized kinetic parameters, respectively, and $p_k =$ ln$\alpha_k$ for $k=1...n_r$. The integrated squared deviation is investigated on the
interval $[z_1 , z_2]$ of the independent variable (time and/or space)~\cite{ZT03}. It has been shown~\cite{VVT85} that $\mathcal{G}(\bP)$ can be
approximated around the nominal parameter set ($\bP^{\textbf{0}}$) as,
\begin{equation}
\label{eq:objec_approx}
\mathcal{G}(\bP) \approx (\Delta\bP)^T \mathbb{S}^T \mathbb{S} (\Delta\bP), 
\end{equation}
\noindent where $\Delta\bP = \bP-\bP^{\textbf{0}}$ and
\begin{equation}
\label{eq:S_tilda}
\mathbb{S}  = \begin{pmatrix} \vardbtilde{S}|_{z_1} \\ \vardbtilde{S}|_{z_2} \\ ... \\ ... \\  \vardbtilde{S}|_{z_m} \end{pmatrix}_{(m. n_{eq})\times n_r}.
\end{equation}

\noindent Here, $\vardbtilde{S}|_{z_i}$ ($i=1,\dots,m$) are normalized sensitivity matrices ($\vardbtilde{S}_{ik}=  \frac{\alpha_k}{\xi_i} \frac{\prt \xi_i}{\prt \alpha_k}$) on a series of $m$ quadrature points on $[z_1,z_2]$ to approximate the integral in Eq.\ (\ref{eq:object_func}). Eigen decomposition of $\mathbb{S}^T \mathbb{S} = A \hat{\Lambda} A^T$ results in:
\begin{equation}
\label{EIG_Q}
\mathcal{G}(\bP) \approx (A^T\Delta\bP)^T  \hat{\Lambda} (A^T\Delta\bP),
\end{equation}
\noindent where $\hat{\Lambda}$ = diag $[ \hat{\lambda}_1, \hat{\lambda}_2, ... \hat{\lambda}_{n_r}]$ is a diagonal matrix containing eigenvalues of $\mathbb{S}^T \mathbb{S}$ (which are real and positive) in descending order ($\hat{\lambda}_1 \geq \hat{\lambda}_2 \geq ... \geq \hat{\lambda}_{n_r}$), and $A=[\baa_1, \baa_2, ... \baa_{n_r}]$ contains the eigenvectors of $\mathbb{S}^T \mathbb{S}$ sorted from left to right with the same order in $\hat{\Lambda}$. Then PCA uses two thresholds ($\hat{\lambda}_{\epsilon}$ \& $a_{\epsilon}$) to select first $j$ sets of reactions ($\baa_1, \baa_2, ... , \baa_j$) who satisfy $\hat{\lambda}_j > \hat{\lambda}_{\epsilon}$ condition and choose every $i$th reaction in each $\baa_j$ set with $|a_{ij}|> a_{\epsilon}$ condition~\cite{EC11}.

\subsection{Sensitivity analysis with optimally time dependent modes}
\label{subsection:SA_OTD_section}

Like PCA our kinetic model reduction strategy is based on selecting reactions, whose perturbations grow  most intensely in the composition evolution given by Eq.\ (\ref{eq:baseODE}). However, the decision about selecting important reactions is made here based on \textit{instantaneous} observation of \textit{modeled} sensitivities in \textit{compressed} format, unlike PCA. In \S~\ref{subsubsection:f_OTD_formulation} we describe the f-OTD methodology for modeling the sensitivity matrix and in \S~\ref{subsubsection:Reaction_selec_f-OTD} we explain our strategy for reaction and species reduction.

\subsubsection{Modeling the sensitivity matrix}
\label{subsubsection:f_OTD_formulation}

Imagine we perturb the composition evolution equation (Eq.\ (\ref{eq:baseODE})) by infinitesimal variations of $\alpha_k=1$ to $\alpha_k=1+\delta \alpha_k$, where $\delta \alpha_k \ll 1$ for $k=1,2,\dots, n_r$. In f-OTD method we expand the sensitivity matrix $S(t)$ into a time-dependent subspace in the $n_{eq}$-dimensional phase space of  compositions represented by a set of f-OTD modes: $U(t) = [\bu_1(t), \bu_2(t), \dots, \bu_r(t)] \in \mathbb{R}^{n_{eq} \times r}$. These modes are orthonormal $\bu_i^T(t) \bu_j(t) = \delta_{ij}$ for any $t$ where $\delta_{ij}$ is the Kronecker delta. The rank of $S(t)\in \mathbb{R}^{n_{eq} \times n_r}$ is $d=\mbox{min}\{n_{eq},n_r\}$ while the f-OTD modes represent a rank-$r$ subspace, where $r\ll d$. In the following we present closed-form evolution equation of the f-OTD modes. To this end, we approximate the sensitivity matrix via the f-OTD decomposition
\begin{equation}
\label{eq:SUY}
S(t) \cong U(t)Y^T(t),
\end{equation}
\noindent where $Y(t) = [\by_1(t), \by_2(t), \dots, \by_r(t)] \in \mathbb{R}^{n_r \times r}$ is the f-OTD coefficient matrix. The above decomposition is not exact as the Eq.\ (\ref{eq:SUY}) is a low-rank approximation of the sensitivity matrix $S(t)$.  Note that in the above decomposition both $U(t)$ and $Y(t)$ are time dependent. We drop the explicit time dependence on $t$ for brevity. Figure~\ref{FIG:S_matrix_shape} shows a schematic of decomposition of $S$ into f-OTD components $U$ and $Y$. The evolution equation for $U$ and $Y$ are obtained by substituting Eq.\ (\ref{eq:SUY}) into Eq.\ (\ref{eq:sensit_exact})
\begin{equation}
\label{eq:sensit_model}
\frac{dS}{dt}  \cong \frac{dU}{dt}Y^T + U\frac{dY^T}{dt} = LUY^T + F.
\end{equation}
\noindent Projecting Eq.\ (\ref{eq:sensit_model}) to $U$ results in
\begin{equation}
\label{eq:sensit_model_proj}
 U^T \frac{dU}{dt}Y^T + U^T U\frac{dY^T}{dt} = U^TLUY^T + U^T F.
\end{equation}
\noindent Since the f-OTD modes are orthonormal, $U^T U =I$. Taking a time derivative of the orthonormality condition results in:
$\dfrac{dU^T}{dt} U + U^T\dfrac{dU}{dt} = 0$. This means that $\varphi= U^T\dfrac{dU}{dt} \in \mathbb{R}^{r\times r}$ is a skew-symmetric matrix ($\varphi^T=-\varphi$). As it was shown in Refs. \cite{BS16,DCB20}, any skew-symmetric choice of matrix $\varphi$ will lead to equivalent f-OTD subspaces. Here we choose $\varphi=0$. Using $U^T U =I$ and  $U^T \dfrac{dU}{dt}=0$, Eq. (\ref{eq:sensit_model_proj}) simplifies to
\begin{equation}
\label{eq:Y_evolutionT}
 \frac{dY^T}{dt} = U^TLUY^T + U^T F.
\end{equation}
\noindent The evolution equation for $U$ can be obtained by substituting $\dfrac{dY^T}{dt}$ from Eq.\ (\ref{eq:Y_evolutionT}) in Eq.\ (\ref{eq:sensit_model}) and  projecting the resulting equation onto $Y$ by multiplying $Y$ from right
\begin{equation}
\label{eq:U_evolution}
\frac{dU}{dt} = QLU+QFYC^{-1},
\end{equation}
\noindent where $Q= I-UU^{T}$ is the orthogonal projection onto the space spanned by the complement of $U$ and $C=Y^TY \in \mathbb{R}^{r \times r}$ is a \emph{correlation matrix} matrix. Matrix $C(t)$ is, in general, a full matrix implying that the f-OTD coefficients are correlated. Equation (\ref{eq:Y_evolutionT}) can be written as
\begin{equation}
\label{eq:Y_evolution}
 \frac{dY}{dt} = Y L_r^T +  F^T U,
\end{equation}
where $L_r = U^T L U \in \mathbb{R}^{r \times r}$ is a reduced linearized operator. Equations (\ref{eq:U_evolution}) and (\ref{eq:Y_evolution}) are a coupled system of ODEs and they constitute the f-OTD evolution equations. The f-OTD modes align themselves with the most instantaneously sensitive directions of the composition evolution equation when perturbed by $\balpha$. It is shown in Ref.\ \cite{BFHS17} that when $\balpha$ is the perturbation to the initial condition, the f-OTD modes converge exponentially fast to the eigen-directions of the Cauchy–Green tensor associated with the most intense finite-time instabilities.   

\sloppy Normalized sensitivity matrix $\vardbtilde{S}(t)$ can be modeled as $\vardbtilde{S}=\Gamma UY^T$ at each time where $\Gamma =$ diag$[1/\xi_1,1/\xi_2, ..., 1/\xi_{n_{eq}}]$. However this matrix is never reconstructed at any time during skeletal model reduction  calculations and we only look for eigen decomposition of $\vardbtilde{S}\vardbtilde{S}^T = (\Gamma U) C (U^T \Gamma)$ as $\vardbtilde{S}\vardbtilde{S}^T = B\hat{\Lambda} B^T$. Here, $\hat{\Lambda}$ = diag $[ \hat{\lambda}_1, \hat{\lambda}_2, ... \hat{\lambda}_{n_s}]$ is a diagonal matrix containing eigenvalues of $\vardbtilde{S}\vardbtilde{S}^T$ (which are real and positive) in descending order ($\hat{\lambda}_1 \geq \hat{\lambda}_2 \geq ... \geq \hat{\lambda}_{n_s}$), and $B=[\bbb_1, \bbb_2, ... \bbb_{n_s}]$ contains the eigenvectors of $\vardbtilde{S}\vardbtilde{S}^T$  sorted from left to right with the same order in $\hat{\Lambda}$. It should be highlighted that $\hat{\Lambda}$ calculated by PCA and f-OTD are comparable; however, $\hat{\Lambda}$ matrix is time dependent in f-OTD calculations while it was defined as time-averaged in PCA. The singular value decomposition of $\vardbtilde{S}$ would be  $\vardbtilde{S} = B \hat{\Sigma} V^T$ where $\Sigma = \hat{\Lambda}^{1/2}$ and $V = Y U^T \Gamma B \Sigma^{-T}$. $V = [\bv_1,\bv_2, ..., \bv_r]$ contains the right singular vectors of $\vardbtilde{S}$ and would be observed for model reduction in next sections.



\subsubsection{Selecting important reactions \& species}
\label{subsubsection:Reaction_selec_f-OTD}

Here is our skeletal model reduction strategy:

\begin{enumerate}
\item Modeled sensitivities are computed in compressed format by solving Eqs.\ (\ref{eq:U_evolution}) and (\ref{eq:Y_evolution}). These two equations are evolved in addition with Eq.\ (\ref{eq:baseODE}) and $\bxi$, $U$, and $Y$ are stored at resolved time steps $t_i \in [0,t_f]$. Equation\  (\ref{eq:baseODE}) is initialized with a set of initial conditions  \textit{i.e.} combination of initial temperatures, equivalence ratios, \textit{etc}. Each simulation case with a different initial condition is called a \textit{sample} here. Equations\ (\ref{eq:U_evolution}) and (\ref{eq:Y_evolution}) are initialized by first solving the SE (Eq.\ (\ref{eq:sensit_exact})) for a few time steps and then performing singular value decomposition of the sensitivity matrix $S= B' \Sigma' {V'}^T$ and assigning the left singular vectors to $U$ and  $Y=V'\Sigma'$.

\item At each resolved time step and for each sample we compute the eigenvalue decomposition  of $\vardbtilde{S} \vardbtilde{S}^T \in \mathbb{R}^{n_s \times n_s}$ where $\vardbtilde{S} = \Gamma UY^T$.
We can eliminate rows and columns of  $\Gamma$ and $U$ matrices associated with species whose mass fractions are negligible (\textit{e.g.} $\xi_i <$1e-8) as we cannot define normalized sensitivities $\vardbtilde{S}_{ik}=  \frac{\alpha_k}{\xi_i} \frac{\prt \xi_i}{\prt \alpha_k}$ for such species. This approach results in taking the eigen decomposition of a matrix with a size of $\mathcal{O}(10)$ at each time location. The other option is to use semi-normalized sensitivities ($\tilde{S}_{ik}=  {\alpha_k} \frac{\prt \xi_i}{\prt \alpha_k}$) instead of normalized sensitivities. Here we use the first approach. 

\item \sloppy Based on Algorithm\ \ref{alg:Spec-Reac-Algo}, we define $\bw(t) \in \mathbb{R}^{N_1\times 1}$ vectors for every resolved time-step and for every sample case. Each element of $\bw(t)$, \textit{i.e.} $w_i$, is positive and associated with a certain reaction. The larger $w_i$ gets, the more important reaction $i$ would be. $\bw(t)$ vectors are then collected to establish $W$ matrix as shown in Fig.\ \ref{fig:W_matrix}. $\bw(t)$ vectors contain the effects of $r$ right singular vectors of $\vardbtilde{S}$ matrix ($\bv_i(t)$) weighted based on their associated eigenvalues of $\vardbtilde{S} \vardbtilde{S}^T$. This prevents us from dealing with each $\bv_i(t)$ separately. As shown in \S3 and \S4, $\hat{\lambda}_1(t)$ is almost always orders of magnitude larger than $\hat{\lambda}_2(t)$ and as the result $\bw(t) \approx |\bv_1(t)|$.

{\centering
\begin{minipage}{.6\linewidth}
\begin{algorithm}[H]
\caption{Sorting reactions based on their importance} \label{alg:Spec-Reac-Algo}
\hspace*{\algorithmicindent} \textbf{Output:} $\bchi$ and $I$ 
\begin{algorithmic}[1]
\Procedure{}{}
\For {all samples}
\For {all resolved time-steps}
\State $\hat{\Lambda} \leftarrow$  first sorted $r$ eigenvalues of $\vardbtilde{S} \vardbtilde{S}^T$;
\State $\hat{V}  \leftarrow$  first sorted right singular vectors of $\vardbtilde{S}$; 
\State Store $\bw =$  $ ( \Sigma^r \hat{\lambda}_i |\bv_i|)/( \Sigma^r \hat{\lambda}_i) \in \mathbb{R}^{n_r}$
\EndFor
\EndFor
\State Construct $W$ as shown in Fig.\ \ref{fig:W_matrix}.
\State [$\bchi$,$I$] = sort(max(abs(W),[],2));
\EndProcedure
\end{algorithmic}
\end{algorithm}
\end{minipage}
\par
}

\item The outputs of Algorithm\ \ref{alg:Spec-Reac-Algo} are $\bchi$ and $I$ which contain sorted measures and indexes for the reactions in the detailed model based on their importance in test cases (ignition here).

\begin{figure}[!htbp]
\centering
 \epsfig{file=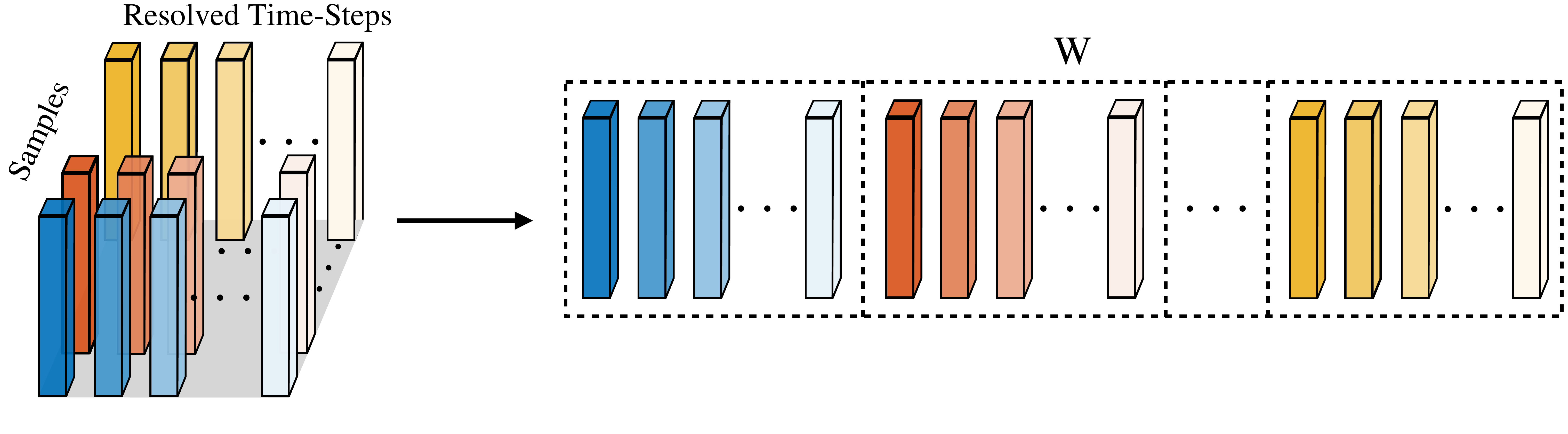,width=14.0cm}
 \caption{Constructing $W$ matrix by collecting $\bw$ vectors at resolved times-steps for every sample. Pillars are representative of $\bw$s.}
\label{fig:W_matrix}
\end{figure}

\item Species are then sorted based on their first presence in ranked reactions, \textit{i.e.} species who first show up  in a higher ranked reaction would the more important than a species who first participate in a lower ranked reaction. As a result we  have a species ranking based on $\bchi$ vector. 

\item Finally, we choose a set of species $\mathbb{M}$ by putting a threshold $\chi_{\epsilon}$ on $\bchi$ vector and eliminate unimportant species and the reactions which they participate in them.


\end{enumerate}


\section{Model reduction with f-OTD: application for hydrogen-oxygen combustion}
\label{section:H2O2_combust}

In this section, the process of eliminating unimportant reactions and species from a  detailed kinetic model with f-OTD is described and its differences with PCA are highlighted. Burke model~\cite{Burke12} for hydrogen-oxygen system which contains $n_s=8$ species\footnote{This kinetic model has 13 species but only 8 species participate in reactions.} and $n_r=54$ irreversible (27 reversible) reactions is considered as the detailed model here. The reduction process is performed by analyzing the ignition phenomenon of an adiabatic, stoichiometric hydrogen-oxygen mixture at atmospheric pressure and $T_0$=1200K with integration of both the SE (Eq.\ (\ref{eq:sensit_exact})) and f-OTD  equations (Eqs.\ (\ref{eq:U_evolution}) and (\ref{eq:Y_evolution})). Exact sensitivities from SE are computed for two purposes, i) finding PCA eigenmodes and eigenvalues, ii) analyzing the performance of f-OTD by comparing the instant eigenvalues of  $\vardbtilde{S} \vardbtilde{S}^T$ at each $t$ both from f-OTD against those obtained by solving the SE. The latter is equivalent to performing instantaneous PCA (I-PCA) on the full sensitivity matrix. The I-PCA shows \emph{the} optimal reduction of the time-dependent sensitivity matrix and we show that the eigenvalues of f-OTD closely approximates the $r$ most dominant eigenvalues of I-PCA.

\begin{figure}[!h]
\centering
 \epsfig{file=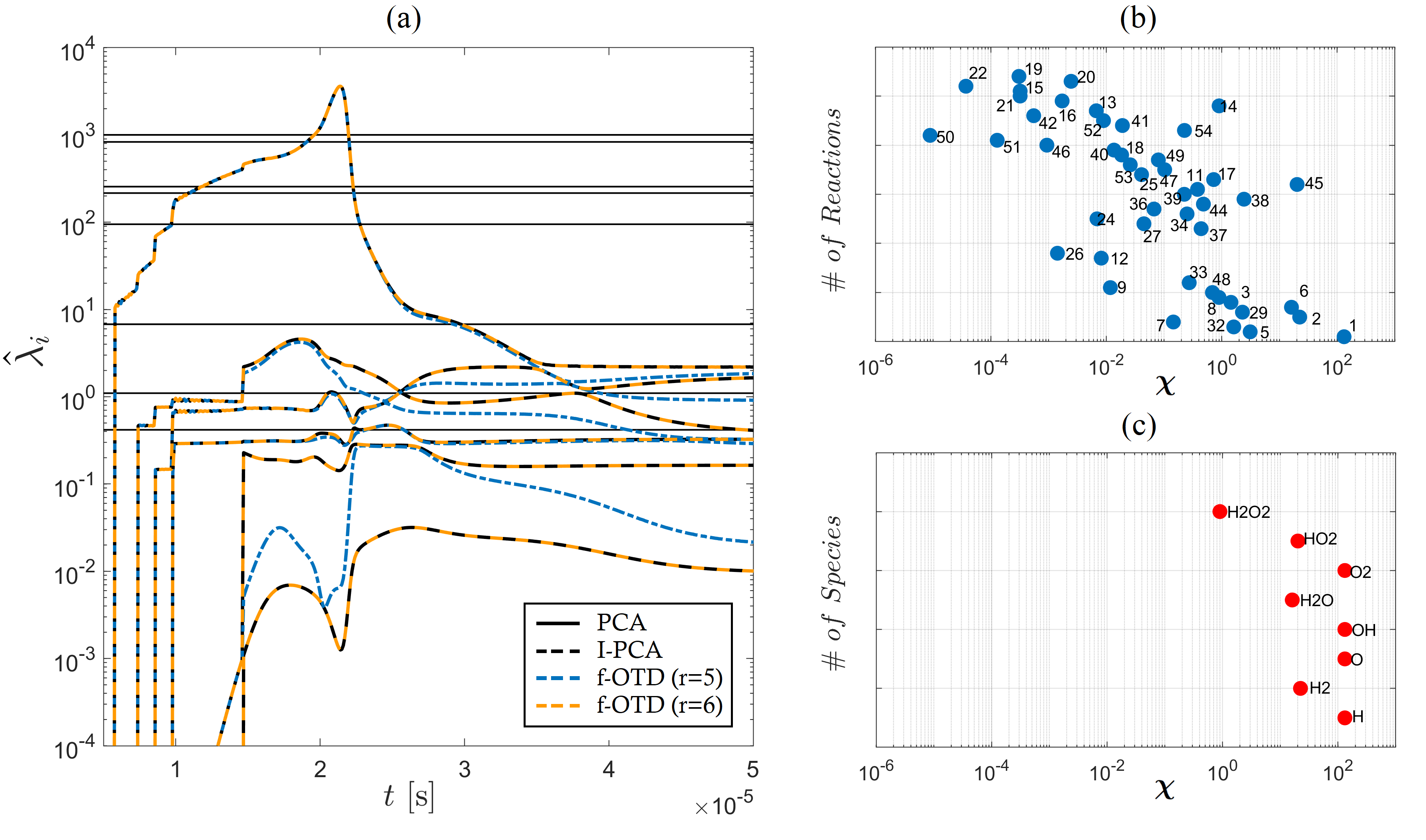,width=16.0cm}
 \caption{Model reduction for hydrogen-oxygen: (a) eigenvalues calculated by PCA, I-PCA and f-OTD, (b,c) sorted reactions and species based on their associated $\chi$ values according to Algorithm\ \ref{alg:Spec-Reac-Algo}. Ignition data is gathered for stoichiometric atmospheric hydrogen-oxygen system with $T_0$=1200K.} 
\label{fig:Sigm_Burke}
\end{figure}
\begin{figure}[!h]
\centering
 \epsfig{file=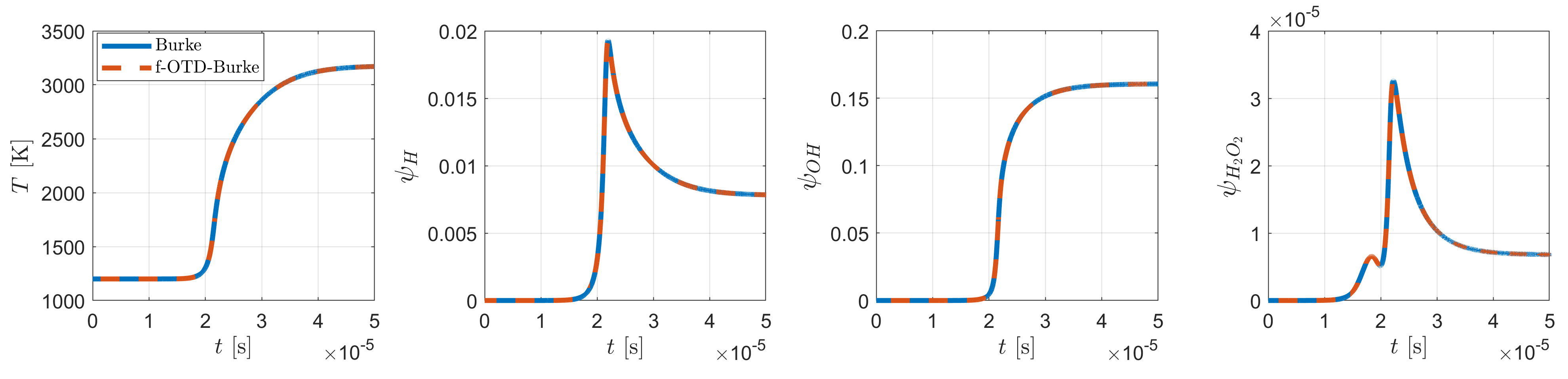,width=16.0cm}
 \caption{Model reduction for hydrogen-oxygen: comparison of the predicted temperature and some species mass fraction profiles from different models with the same conditions shown in Fig.\ \ref{fig:Sigm_Burke}.}
\label{fig:Burke_PsiT}
\end{figure}

Figure\ \ref{fig:Sigm_Burke}(a) compares top eigenvalues of f-OTD with PCA (static) and I-PCA (instantaneous). It is shown that the top PCA eigenvalues are time invariant and close to each other. In contrast the first f-OTD eigenvalue is orders of magnitude larger than the others during the course of ignition, \textit{i.e.} from $t$=0 to $t$=30 $\mu$s (until most of the heat is released). Moreover, f-OTD eigenvalues match with I-PCA with increasing number of modes ($r$). This means that the modeled sensitivities converge to the exact values by adding more modes, in this case addition of top 6 modes. The results also show that with f-OTD (r=5), the time variation of top eigenvalues is captured well while the second dominant eigenvalues deviates from I-PCA solution in the main non-equilibrium reaction layer and the post heat release region. Figure\ \ref{fig:Sigm_Burke}(b) portrays the species ranking generated by Algorithm\ \ref{alg:Spec-Reac-Algo} versus $\bchi$. 
Figure~\ref{fig:Burke_PsiT} demonstrates f-OTD-Burke ability in reproducing the species evolution using the Burke model, for a stoichiometric mixture of  hydrogen-oxygen at $p$=1 atm and $T_0$=1200K.

\begin{figure}[!h]
\centering
 \epsfig{file=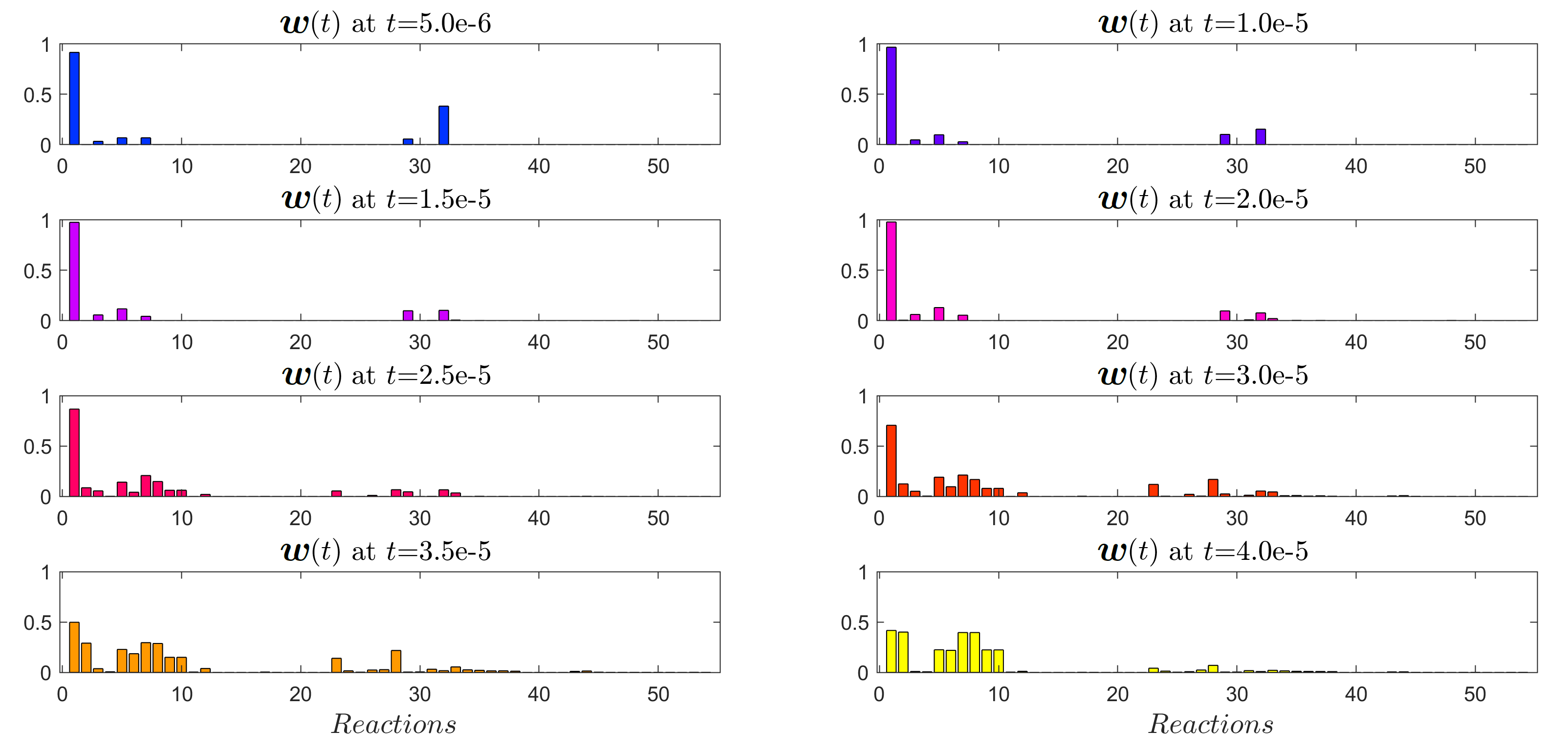,width=16.0cm}
 \caption{Model reduction for hydrogen-oxygen: temporal variation of $\bw(t)$ in  stoichiometric atmospheric hydrogen-oxygen ignition simulation with $T_0=$1200K.}
\label{fig:Burke_OTD_modes}
\end{figure}

Figure\ \ref{fig:Burke_OTD_modes} portrays the temporal evolution of $\bw(t)$. As each element of $\bw(t)$ is associated with a reaction, every change in the shape of this temporal vector signifies a change in the importance of reactions during the course of ignition. For example the first reaction (H+O$_2$ $\rightarrow$ O+OH) is the most important one at $t$=5.0e-6 but as marching in time and approaching post peak heat release region, other reactions (\textit{e.g.} reaction\# 2-10, specifically radical recombination reactions like OH+OH$\rightarrow$O+H$_2$O also become important. Algorithm\ \ref{alg:Spec-Reac-Algo} ranks reactions based on their maximum value on $\bw$ in different times and samples. This demonstration shows that f-OTD detects  key reactions that become important only for a short period of time. Such reactions could not be identified and selected for removal in static model reduction methods such as PCA where ranking is performed in a time-averaged sense unless a large number of modes are retained in the reduced representation  \cite{EC11}.

\begin{figure}[!htbp]
\centering
 \epsfig{file=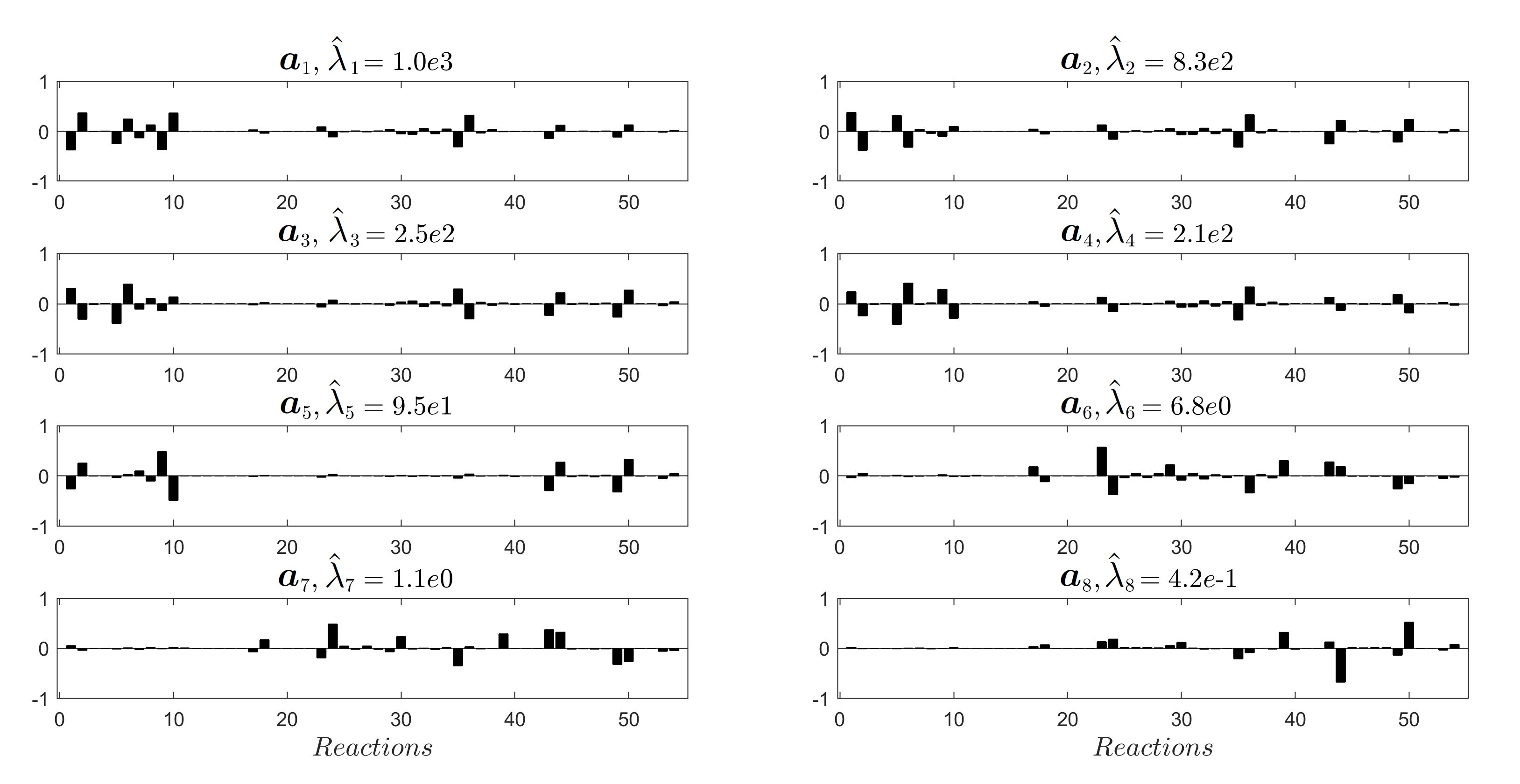,width=16.0cm}
 \caption{Model reduction for hydrogen-oxygen: first eight eigenmodes of $\mathbb{S}^T\mathbb{S}$ calculated by PCA, from combined analysis over the time interval 0 to 5.0e-6 secs.}
\label{fig:Burke_PCA_modes}
\end{figure}

Figure~\ref{fig:Burke_PCA_modes} shows the first 8 eigenmodes of $\mathbb{S}^T\mathbb{S}$ matrix (Eq.\ (\ref{eq:S_tilda})) and their associated eigenvalues with PCA. $\hat{\lambda}_1$-$\hat{\lambda}_4$ have a same order of magnitude and $\baa_1$-$\baa_4$ introduce similar reactions as important ones. Reaction \# 22 and 28 are not significant reactions in $\baa_1$-$\baa_4$. However, it was shown in Fig.\ \ref{fig:Burke_OTD_modes} that these reactions are effective in $t \in$[3.0e-5,3.5e-5]. The next step in model reduction with PCA is to define thresholds $\hat{\lambda}_{\epsilon}$ and $a_{\epsilon}$ to select important reactions as described in \S~\ref{subsection:PCA_sec}. This process is beyond the scope of this study.

\section{Skeletal reduction: application for ethylene-air burning}
\label{section:Skeletal}

\begin{figure}[!h]
\centering
 \epsfig{file=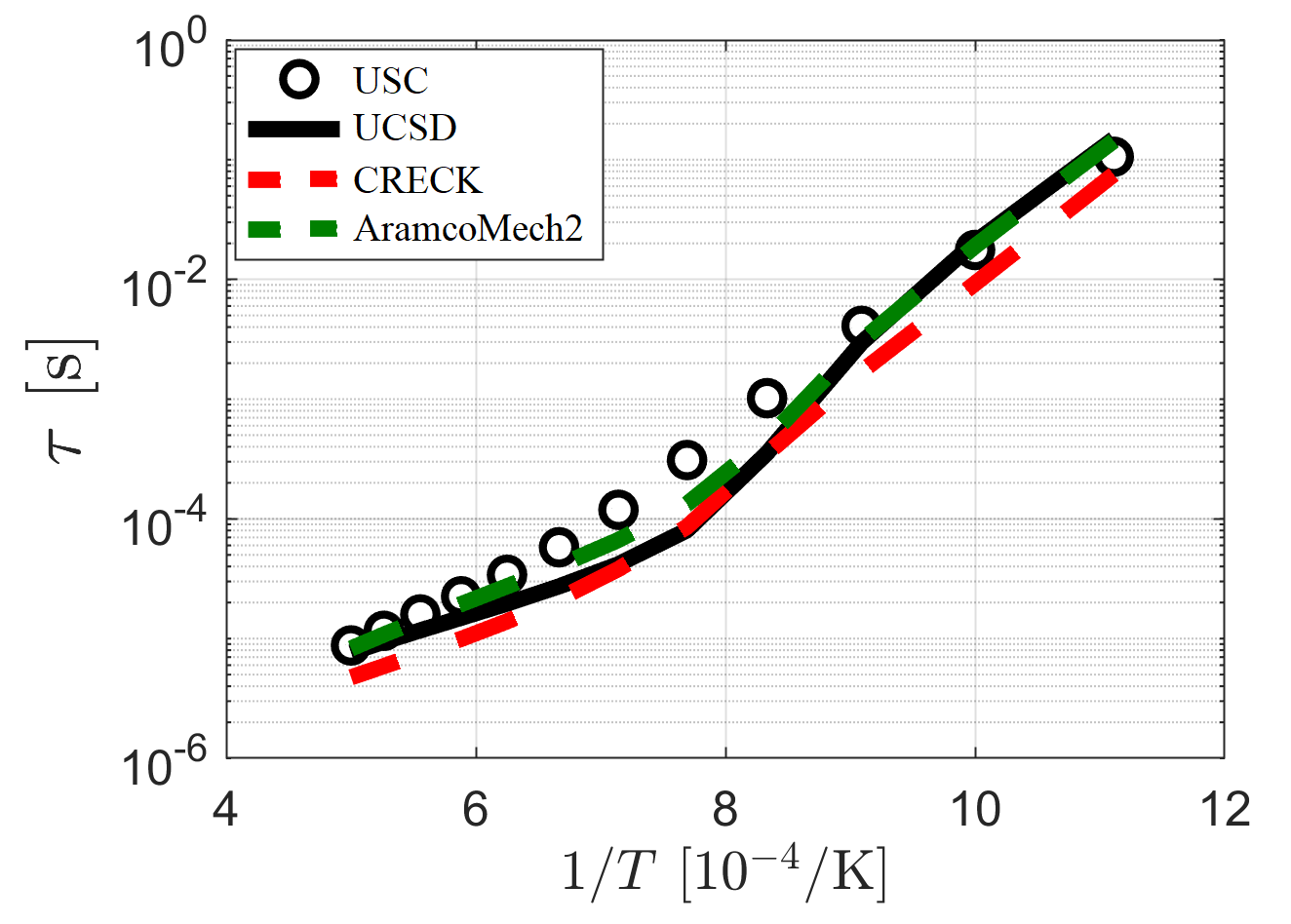,width=8.0cm}
 \caption{Skeletal model reduction for ethylene-air: ignition delays calculated by different detailed kinetic models for an atmospheric stoichiometric mixture of ethylene-air. USC, UCSD, CRECK, and AramcoMech2 are in good agreement with each other.}
\label{fig:detailed_models}
\end{figure}

There are several detailed kinetic models for ethylene–air burning in literature developed at different institutions, \textit{e.g.}  at the University of California, San Diego (UCSD)~\cite{UCSD_EthylAir}, the University of Southern California (USC - a subset of JetSurf)~\cite{SYWL09}, the KAUST  (AramcoMech2)~\cite{SAM}, and the Politecnico of Milan  (CRECK)~\cite{CRECK}. Figure\ \ref{fig:detailed_models} shows the ignition delays as calculated by all of these models  are in a reasonable agreement with each other. Moreover, it is shown in  Ref.~\cite{PPS18}  that USC ignition delays are closer to experimental data in comparison to  those by the USCD, CRECK, and AramcoMech2 models. Therefore, here we use USC as our detailed kinetic model, and extract a series of skeletal models from this baseline.

\subsection{Problem setup and initial conditions}
\label{subsection:Skeletal_init}

Simulations are conducted of an adiabatic, atmospheric pressure reactor with different initial temperatures $T_0 \in $[1400,2000] and equivalence ratios $\phi \in $[0.5,1.0,1.5] for ethylene-air mixture. The USC model~\cite{SYWL09} with 111 species and 1566 irreversible (784 reversible) reactions is our detailed model from which all the skeletal models (f-OTDs) are generated. Only three f-OTD modes ($r=3$) are used to model the sensitivity matrix. Simulations with SK31~\cite{EC11}, SK32~\cite{Luo12} and SK38~\cite{EC11}, which are also skeletal models generated from two versions of USC (optimized and unoptimized), are provided here  for comparison.  The comparisons are made based on three criteria: i) ignition delay, ii) premixed flame speed, iii) non-premixed extinction strain, which is the maximum axial velocity gradient on the air-side of axisymmetric opposed jet ethylene-air flame. Flame speeds and extinction curves are generated  by Cantera~\cite{Cantera}.

\subsection{Skeletal models}
\label{subsection:model_reduct_f-OTD}

\begin{figure}[!htbp]
\centering
 \epsfig{file=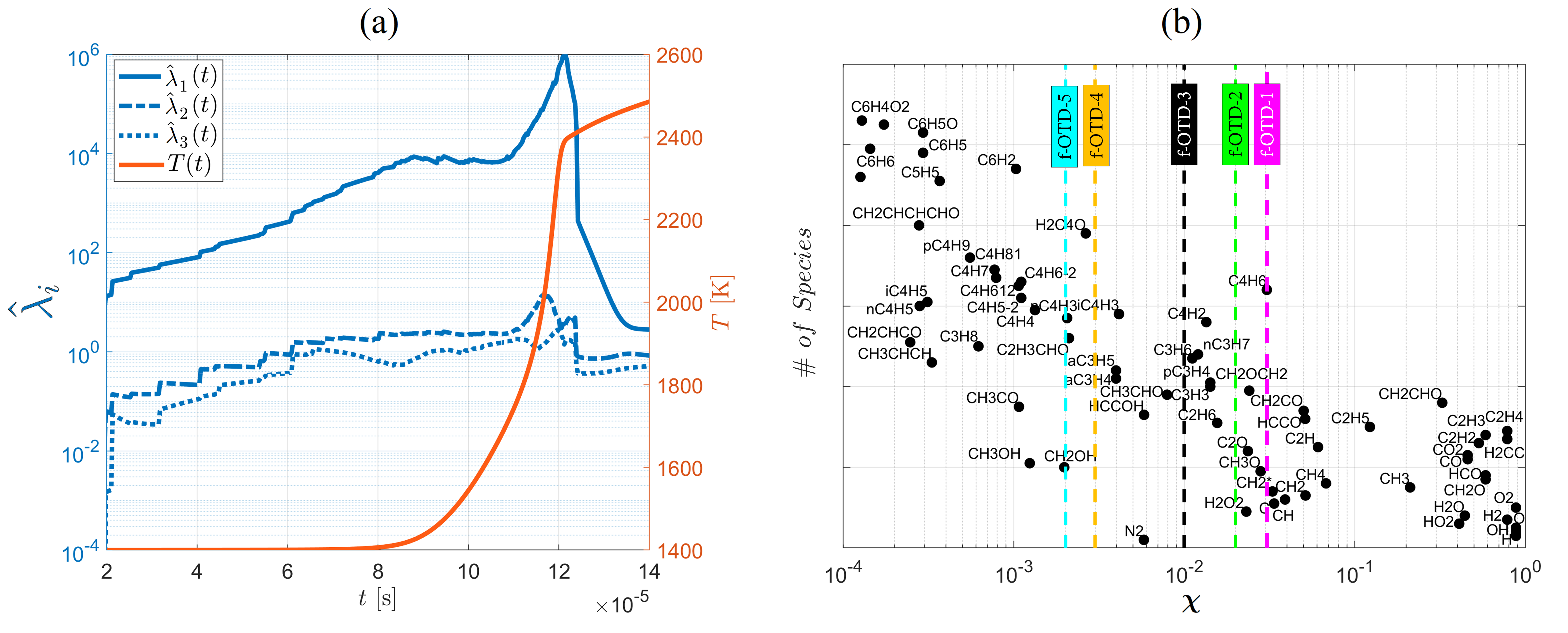,width=16.0cm}
 \caption{Skeletal model reduction for ethylene-air: (a) eigenvalues of $\vardbtilde{S} \vardbtilde{S}^T$ with $r=3$ for ignition simulation initialized with $T_0$=1200K and $\phi$=1.0. $\hat{\lambda}_1(t)$ is orders of magnitude larger than the others during ignition. (b) Species ranking based Algorithm\ \ref{alg:Spec-Reac-Algo} with $\chi_{\epsilon}$ associated with f-OTD models. Only species with $\chi >$1.0e-4 are presented in this figure.}
\label{fig:Spec_rank}
\end{figure}

Figure\ \ref{fig:Spec_rank}(a) portrays the evolution of eigenvalues of $\vardbtilde{S} \vardbtilde{S}^T$ in time. $\hat{\lambda}_1(t)$ is two orders of magnitude larger than $\hat{\lambda}_2(t)$ during ignition except around temperature inflection point\footnote{Temperature at $dT/dt|_{max}$} ($T_{inf}$) in which $\hat{\lambda}_1(t)$ is six orders of magnitude larger than $\hat{\lambda}_2(t)$. This means that only one mode is dominant during ignition and contain more than 95\% of the energy of the dynamical system ($\hat{\lambda}_1(t) / \Sigma_{i=1}^r\hat{\lambda}_i(t) > 0.95$). Figure\ \ref{fig:Spec_rank}(b) shows the species ranking based on the process described in Algorithm\ \ref{alg:Spec-Reac-Algo}. Similar to the rankings in H$_2$-O$_2$ system presented in \S\ref{section:H2O2_combust}, the species H, OH, O and O$_2$ associated with the most sensitive reaction H+O$_2 \rightarrow$ OH+O appear as the most important species based on their associated values on $\bchi$ vector. Different reduced order models can be generated by varying the threshold $\chi_{\epsilon}$ on $\bchi$ and eliminating those species with $\chi<\chi_{\epsilon}$ and associated reactions they participate in. Our goal is to find a model which can reproduce the results of USC model based on the criteria mentioned in \S\ref{subsection:Skeletal_init} with a pre-determined accuracy, \textit{e.g.} less than 1\% error.  Table\ \ref{table:models} provides the details of models generated with varying threshold, $\chi_{\epsilon}$.

\begin{table}[!ht]
\caption{Model Characteristics}\label{table:models}
\scriptsize
\centering

\begin{tabular}{M{2.0cm}M{1.0cm}M{1.0cm}M{1.0cm}} \hline   \toprule   Model &  $\chi_{\epsilon}$ & $n_s$  & $n_r$  \\ \hline \hline
USC  & - & 111  & 1566  \\ \hline
SK38 & -& 38  & 474 \\ \hline
SK32 & -& 32  & 412 \\ \hline
f-OTD-1 & 3e-2 & 28 & 324 \\ \hline
f-OTD-2 & 2e-2 & 32  & 386 \\ \hline
f-OTD-3 & 1e-2 & 38  & 472 \\ \hline
f-OTD-4 & 3e-3 & 43  & 570 \\ \hline
f-OTD-5 & 2e-3 & 46  & 610 \\ \hline
\end{tabular}
\end{table}

\begin{figure}[!h]
\centering
 \epsfig{file=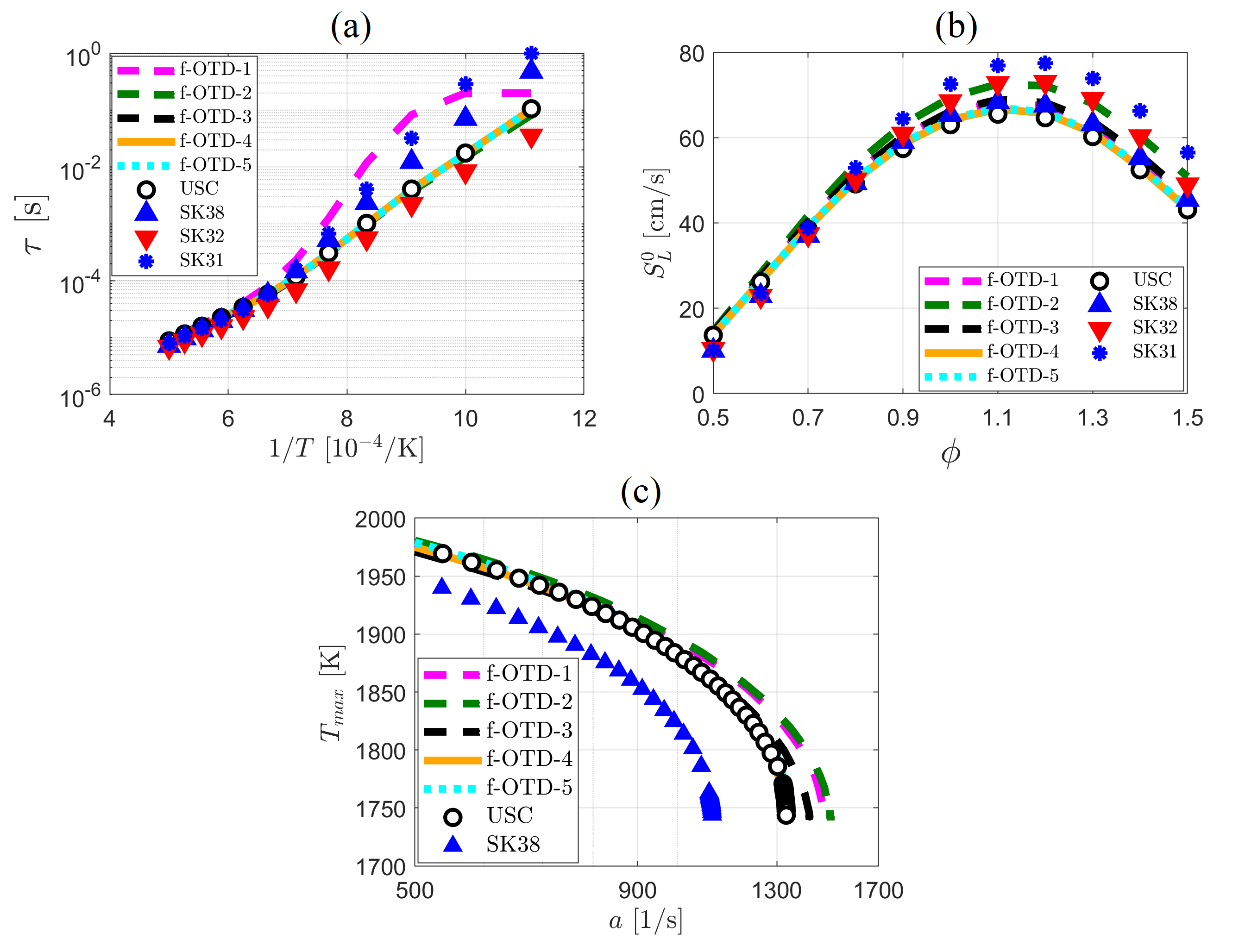,width=14.0cm}
 \caption{Skeletal model reduction for ethylene-air: (a) predicted ignition delays, (b) flame speeds, and (c) extinction strains for skeletal models generated from USC. Ignition delays  are generated  with $\phi_0$=1.0, flame speeds with $T_0$=300K, both at 1atm pressure. Extinction curves are generated for ethylene-air diffusion flame at 1atm pressure and $T_0$=300K.}
\label{fig:Ign_Sl_Ext}
\end{figure}

Figure\ \ref{fig:Ign_Sl_Ext}(a) demonstrates that f-OTD models with $n_s \ge 32$ perfectly estimate the ignition delays for a stoichiometric mixture of ethylene-air. SK32 under-predicts while SK38 and SK31 (with slightly different rate constants from USC) over-predict the ignition delays. Figure\ \ref{fig:Ign_Sl_Ext}(b) shows that the UCSD ignition delay predictions lies between the predictions of USC and CRECK models. Figure\ \ref{fig:Ign_Sl_Ext}(c) compares the laminar premixed flame speeds predicted by different models initialized with $T_0$=300K. f-OTD-2, SK31 and SK32 have largest differences from USC model. f-OTD models with $n_s \ge 38$ and SK38 have the best flame speed predictions even tough f-OTD models show better results at lower and upper bounds of $\phi_0$. Figure\ \ref{fig:Ign_Sl_Ext}(e) compares the extinction strain rate as predicted by our models. All f-OTD models show good agreement in estimating extinction strain rates, which is the toughest canonical flame feature to predict. SK38 under-predicts the maximum temperatures indicating the influence of optimized rate constants in Ref.~\cite{SYWL09}.

\begin{figure}[!h]
\centering
 \epsfig{file=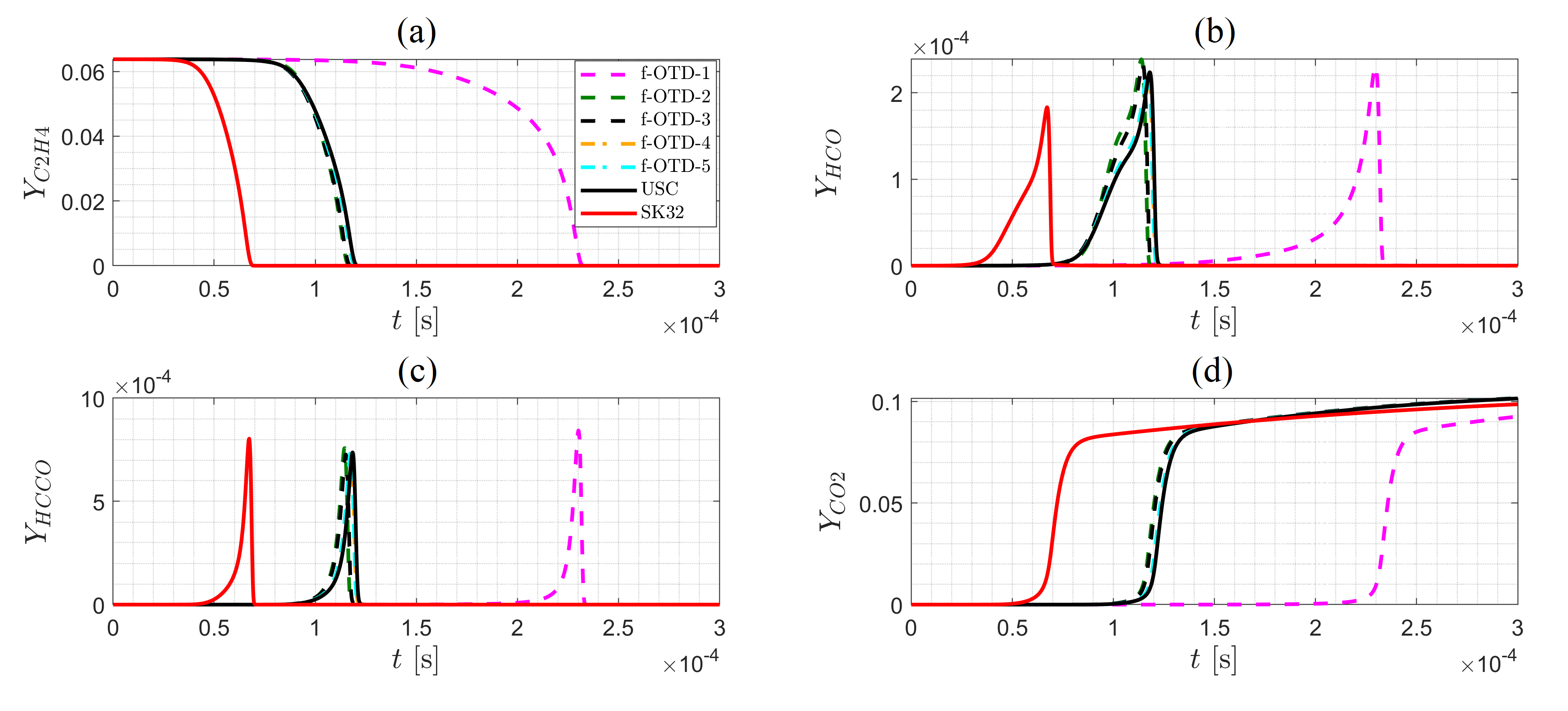,width=16.0cm}
 \caption{Skeletal model reduction for ethylene-air: Evolution of (a) $C2H4$, (b) $HCO$, (c) $HCCO$, and (d) $CO2$ mass fractions as predicted by different models in Table\ \ref{table:models} with $T_0$=1400K, $\phi$=1.0, and at 1atm pressure. f-OTD models with $n_s \ge 32$ show strong ability in reproducing USC results.}
\label{fig:Spec_Mass}
\end{figure}

Figure\ \ref{fig:Spec_Mass} portrays the species mass fraction evolution for some key species in a mixture initialized with $\phi_0$=1.0 and $T_0$=1400K. This figure highlights the ability of f-OTD-2 model with 32 species in predicting the ignition phenomenon. Moreover, all f-OTD models (with $n_s \ge 32$) provide a better estimate for the maximum mass fraction of species shown in Fig.\ \ref{fig:Spec_Mass} comparing with SK32, similar to conclusions drawn in ref.~\cite{EC11}.

\begin{figure}[!h]
\centering
 \epsfig{file=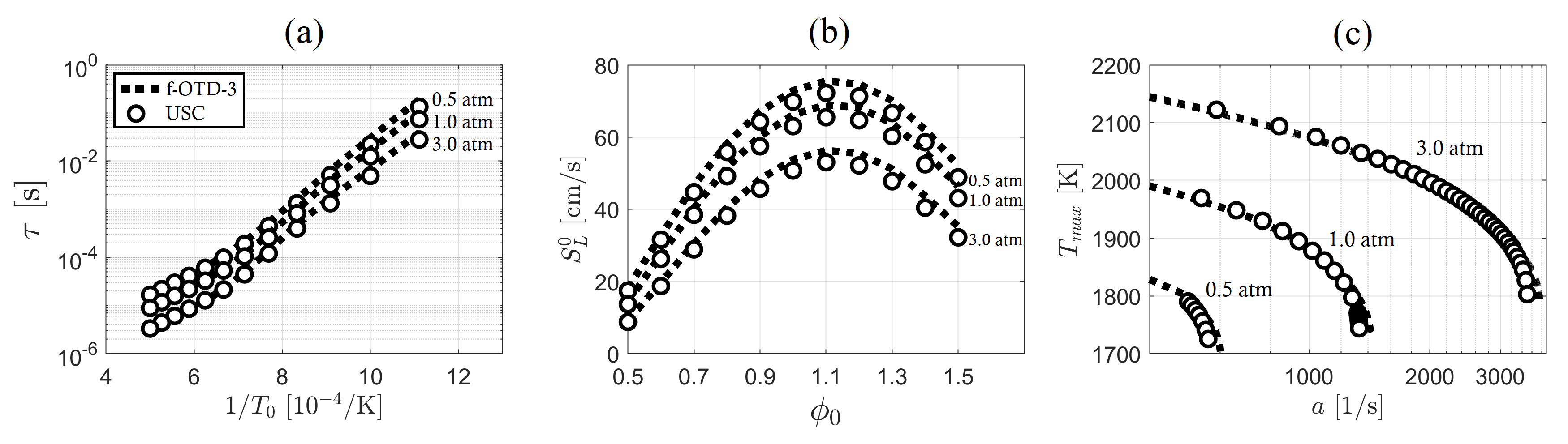,width=16.0cm}
 \caption{Skeletal model reduction for ethylene-air: comparison of (a) ignition delay, (b) flame speed, and (c)  extinction strain as predicted by the detailed model (USC) and the f-OTD-3 skeletal model. Ignition delays are generated for $\phi_0$=1.0, and flame speed and extinction simulations are  performed with $T_0$=300K.}
\label{fig:fOTD3-Press}
\end{figure}

Observing the results provided in Fig.\ \ref{fig:Ign_Sl_Ext}, it is clear that f-OTD-1 is not a good skeletal model for USC. This can be primarily attributed to the elimination of C$_2$O, CH$_2$OCH$_2$, CH$_3$O, and H$_2$O$_2$ in this 28 species model. Although f-OTD-1 cannot predict the ignition delay accurately, it shows reasonably good performance in estimating laminar flame speeds and maximum temperatures for extinction. As mentioned above, f-OTD-2 perfectly estimates the ignition delays and extinction strain rate. It's flame speed predictions are also matches with SK32 and lies within the predictions of UCSD and USC. Predictions of f-OTD-3, f-OTD-4, and f-OTD-5 are so close to USC and these predictions become more precise with increasing $n_s$. Comparing f-OTD-2 to f-OTD-5 based on their application in reproducing USC model results and also their computational cost, we suggest using f-OTD-2 and f-OTD-3 models. f-OTD-2 and SK32 both have 32 species containing 27 species in common. f-OTD-2 has C,  C$_2$H, C$_2$O, C$_4$H$_6$, CH$_2$OCH$_2$ but SK32 has C$_2$H$_6$, C$_3$H$_6$, CH$_3$CHO, aC$_3$H$_5$, nC$_3$H$_7$ instead. f-OTD-3 and SK38 both have 38 species containing 36 species in common. f-OTD-3 contains C$_3$H$_3$ and pC$_3$H$_4$ while SK38 has aC$_3$H$_5$ and iC$_4$H$_3$. Figure\ \ref{fig:fOTD3-Press} demonstrates the ability of f-OTD-3 in predicting ignition delay, laminar flame  speed and  extinction for three different pressures \textit{i.e.} 0.5atm, 1.0atm, and 3.0atm. f-OTD-3 shows strong ability in reproducing USC model results.

\section{Conclusions}
\label{section:conclu}
Sensitivity analysis with f-OTD is described and implemented for skeletal model reduction. A key feature of the present f-OTD approach is that the sensitivity matrix is estimated as a multiplication of two low-ranked time-dependent matrices which  evolve based on evolution equations derived from the governing equations of the system. Modeled sensitivities are then normalized and reactions and species of a detailed model are ranked in a computationally efficient manner based on their importance to any canonical problem. Different skeletal models can be generated by eliminating unimportant species and their associated reactions from the detailed model. The performance of f-OTD for skeletal model reduction is demonstrated for reducing a detailed C$_1-$C$_4$ hydrocarbon kinetic model with ethylene as the fuel of interest. The generated models are compared based on their ability in predicting ignition delays, flame speeds and diffusion flame extinction strain rates. The results also compared against two skeletal and detailed models reported in the literature. f-OTD demonstrates strong ability in eliminating unimportant species and reactions from the detailed model in an efficient manner. The extension of this study would include sensitivity analysis based on the most effective thermochemistry parameters \textit{e.g.} activation energies, formation enthalpies, and transport properties \textit{e.g.} heat and mass diffusivities. Most importantly, as it was shown recently \cite{DCB20},  f-OTD can be used in solving PDEs for multi-dimensional combustion problems in a cost-effective manner --- by exploiting the correlations between the spatiotemporal sensitivities of different species with respect to different parameters. This analysis can be especially insightful for problems containing rare events \textit{e.g.} deflagration-detonation-transition by providing more insight about the global effective phenomena.

\section*{Acknowledgments}
This work has been co-authored by an employee of Triad National Security, LLC which operates Los Alamos National Laboratory under Contract No. 89233218CNA000001 with the U.S. Department of Energy/National Nuclear Security Administration. The work of PG was supported by Los Alamos National Laboratory, under Contract 614709.  Additional support for the work at Pitt with H.B. as the PI is provided by NASA Transformational Tools and Technologies (TTT) Project Grant  80NSSC18M0150, and by  NSF under Grant CBET-2042918.

\typeout{}
\bibliography{ROM_combustion,HB,cfd_arash}

\begin{thebibliography}{10}
\expandafter\ifx\csname url\endcsname\relax
  \def\url#1{\texttt{#1}}\fi
\expandafter\ifx\csname urlprefix\endcsname\relax\def\urlprefix{URL }\fi
\expandafter\ifx\csname href\endcsname\relax
  \def\href#1#2{#2} \def\path#1{#1}\fi

\bibitem{EC11}
G.~Esposito, H.~Chelliah, {S}keletal {R}eaction {M}odels based on {P}rincipal
  {C}omponent {A}nalysis: {A}pplication to {E}thylene--{A}ir {I}gnition,
  {P}ropagation, and {E}xtinction {P}henomena, Combust. Flame 158~(3) (2011)
  477--489.

\bibitem{SYWL09}
D.~A. Sheen, X.~You, H.~Wang, T.~L{\o}v{\aa}s, {S}pectral {U}ncertainty
  {Q}uantification, {P}ropagation and {O}ptimization of a {D}etailed {K}inetic
  {M}odel for {E}thylene {C}ombustion, Proc. Combust. Inst. 32~(1) (2009)
  535--542.

\bibitem{JetSurf}
H.~Wang, E.~Dames, B.~Sirjean, D.~Sheen, R.~Tangko, A.~Violi, J.~Lai,
  F.~Egolfopoulos, D.~Davidson, R.~Hanson, C.~Bowman, C.~Law, W.~Tsang,
  N.~Cernansky, D.~Miller, R.~Lindstedt, {A} {H}igh-temperature {C}hemical
  {K}inetic {M}odel of n-{A}lkane (up to n-{D}odecane), {C}yclohexane, and
  {M}ethyl-, {E}thyl-, n-{P}ropyl and n-{B}utyl-cyclohexane {O}xidation at
  {H}igh {T}emperatures ({J}et{S}urf 2.0), Tech. rep.,
  http://melchior.usc.edu/JetSurF/JetSurF2.0 (2011).

\bibitem{UCSD_EthylAir}
\href{http://combustion.ucsd.edu}{``{C}hemical-{K}inetic {M}echanisms for
  {C}ombustion {A}pplications," {S}an {D}iego {M}echanism {W}eb {P}age,
  {M}echanical and {A}erospace {E}ngineering ({C}ombustion {R}esearch),
  {U}niversity of {C}alifornia at {S}an {D}iego.}
\newline\urlprefix\url{http://combustion.ucsd.edu}

\bibitem{CRECK}
E.~Ranzi, A.~Frassoldati, R.~Grana, A.~Cuoci, T.~Faravelli, A.~Kelley, C.~Law,
  {H}ierarchical and {C}omparative {K}inetic {M}odeling of {L}aminar {F}lame
  {S}peeds of {H}ydrocarbon and {O}xygenated {F}uels, Prog. Energy Combust.
  Sci. 38~(4) (2012) 468--501.

\bibitem{AramcoMech2}
C.~Zhou, Y.~Li, E.~O'connor, K.~Somers, S.~Thion, C.~Keesee, O.~Mathieu,
  E.~Petersen, T.~DeVerter, M.~Oehlschlaeger, et~al., {A} {C}omprehensive
  {E}xperimental and {M}odeling {S}tudy of {I}sobutene {O}xidation, Combust.
  Flame 167 (2016) 353--379.

\bibitem{Smooke91}
M.~D. Smooke, {R}educed {K}inetic {M}echanisms and {A}symptotic
  {A}pproximations for {M}ethane-{A}ir {F}lames: {A} {T}opical {V}olume,
  Springer, 1991.

\bibitem{Peters1993}
N.~Peters, B.~Rogg (Eds.), Reduced Kinetic Mechanisms for Applications in
  Combustion Systems, Vol.~15 of Lecture Notes in Physics, Springer-Verlag,
  Berlin, Germany, 1993.

\bibitem{Turanyi90a}
T.~Turanyi, {R}eduction of {L}arge {R}eaction {M}echanisms, New J. Chem.
  14~(11) (1990) 795--803.

\bibitem{wang:1991}
H.~Wang, M.~Frenklach, {D}etailed {R}eduction of {R}eaction {M}echanisms for
  {F}lame {M}odeling, Combust. Flame 87~(3-4) (1991) 365--370.

\bibitem{SCGJ10}
W.~Sun, Z.~Chen, X.~Gou, Y.~Ju, {A} {P}ath {F}lux {A}nalysis {M}ethod for the
  {R}eduction of {D}etailed {C}hemical {K}inetic {M}echanisms, Combust. Flame
  157~(7) (2010) 1298--1307.

\bibitem{LL05}
T.~Lu, C.~Law, {A} {D}irected {R}elation {G}raph {M}ethod for {M}echanism
  {R}eduction, Proc. Combust. Inst. 30~(1) (2005) 1333--1341.

\bibitem{ZLL07}
X.~Zheng, T.~Lu, C.~Law, {E}xperimental {C}ounterflow {I}gnition {T}emperatures
  and {R}eaction {M}echanisms of 1, 3-{B}utadiene, Proc. Combust. Inst. 31~(1)
  (2007) 367--375.

\bibitem{LL09}
T.~Lu, C.~Law, {T}oward {A}ccommodating {R}ealistic {F}uel {C}hemistry in
  {L}arge-{S}cale {C}omputations, Prog. Energy Combust. Sci. 35~(2) (2009)
  192--215.

\bibitem{NSR10}
K.~Niemeyer, C.~Sung, M.~Raju, {S}keletal {M}echanism {G}eneration for
  {S}urrogate {F}uels using {D}irected {R}elation {G}raph with {E}rror
  {P}ropagation and {S}ensitivity {A}nalysis, Combust. Flame 157~(9) (2010)
  1760--1770.

\bibitem{Turanyi90b}
T.~Tur{\'a}nyi, {S}ensitivity {A}nalysis of {C}omplex {K}inetic {S}ystems.
  {T}ools and {A}pplications, J. Math. Chem. 5~(3) (1990) 203--248.

\bibitem{Saltelli04}
A.~Saltelli, S.~Tarantola, F.~Campolongo, M.~Ratto, {S}ensitivity {A}nalysis in
  {P}ractice: {A} {G}uide to {A}ssessing {S}cientific {M}odels, Vol.~1, Wiley
  Online Library, 2004.

\bibitem{ESC12}
G.~Esposito, B.~Sarnacki, H.~Chelliah, {U}ncertainty {P}ropagation of
  {C}hemical {K}inetics {P}arameters and {B}inary {D}iffusion {C}oefficients in
  {P}redicting {E}xtinction {L}imits of {H}ydrogen/{O}xygen/{N}itrogen
  {N}on-premixed {F}lames, Combust. Theory Model. 16~(6) (2012) 1029--1052.

\bibitem{Shuang16}
S.~Li, B.~Yang, F.~Qi, {A}ccelerate {G}lobal {S}ensitivity {A}nalysis using
  {A}rtificial {N}eural {N}etwork {A}lgorithm: {C}ase {S}tudies for
  {C}ombustion {K}inetic {M}odel, Combust. Flame 168 (2016) 53--64.

\bibitem{Sobol90}
I.~Sobol', {O}n {S}ensitivity {E}stimation for {N}onlinear {M}athematical
  {M}odels, Matematicheskoe modelirovanie 2~(1) (1990) 112--118.

\bibitem{HS96}
T.~Homma, A.~Saltelli, {I}mportance {M}easures in {G}lobal {S}ensitivity
  {A}nalysis of {N}onlinear {M}odels, Reliab. Eng. Syst. Saf. 52~(1) (1996)
  1--17.

\bibitem{RA99}
H.~Rabitz, {\"O}.~Ali{\c{s}}, {G}eneral {F}oundations of {H}igh-{D}imensional
  {M}odel {R}epresentations, J. Math. Chem. 25~(2-3) (1999) 197--233.

\bibitem{Sobol01}
I.~Sobol, {G}lobal {S}ensitivity {I}ndices for {N}onlinear {M}athematical
  {M}odels and their {M}onte {C}arlo {E}stimates, Math. Comput. Simul. 55~(1-3)
  (2001) 271--280.

\bibitem{LSR02}
G.~Li, S.~Wang, H.~Rabitz, {P}ractical {A}pproaches to {C}onstruct {RS-HDMR}
  {C}omponent {F}unctions, J. Phys. Chem. A 106~(37) (2002) 8721--8733.

\bibitem{VVT85}
S.~Vajda, P.~Valko, T.~Turanyi, {P}rincipal {C}omponent {A}nalysis of {K}inetic
  {M}odels, Int. J. Chem. Kinet. 17~(1) (1985) 55--81.

\bibitem{brown:1997}
N.~Brown, G.~Li, M.~Koszykowski, {M}echanism {R}eduction via {P}rincipal
  {C}omponent {A}nalysis, Int. J. Chem. Kinet. 29~(6) (1997) 393--414.

\bibitem{SFCFR16}
A.~Stagni, A.~Frassoldati, A.~Cuoci, T.~Faravelli, E.~Ranzi, {S}keletal
  {M}echanism {R}eduction {T}hrough {S}pecies-{T}argeted {S}ensitivity
  {A}nalysis, Combust. Flame 163 (2016) 382--393.

\bibitem{DCB20}
M.~Donello, M.~Carpenter, H.~Babaee, {C}omputing {S}ensitivities in
  {E}volutionary {S}ystems: {A} {R}eal-{T}ime {R}educed {O}rder {M}odeling
  {S}trategy, arXiv preprint arXiv:2012.14028.

\bibitem{BOR15}
K.~Braman, T.~Oliver, V.~Raman, {A}djoint-based {S}ensitivity {A}nalysis of
  {F}lames, Combust. Theory Model. 19~(1) (2015) 29--56.

\bibitem{LCRPS19}
M.~Lemke, L.~Cai, J.~Reiss, H.~Pitsch, J.~Sesterhenn, {A}djoint-based
  {S}ensitivity {A}nalysis of {Q}uantities of {I}nterest of {C}omplex
  {C}ombustion {M}odels, Combust. Theory Model. 23~(1) (2019) 180--196.

\bibitem{Langer2020}
R.~Langer, J.~Lotz, L.~Cai, F.~vom Lehn, K.~Leppkes, U.~Naumann, H.~Pitsch,
  {A}djoint {S}ensitivity {A}nalysis of {K}inetic, {T}hermochemical, and
  {T}ransport {D}ata of {N}itrogen and {A}mmonia {C}hemistry, Proc. Combust.
  Inst.

\bibitem{BS16}
H.~Babaee, T.~Sapsis, {A} {M}inimization {P}rinciple for the {D}escription of
  {M}odes {A}ssociated with {F}inite-{T}ime {I}nstabilities, Proc. R. Soc. A
  472~(2186) (2016) 20150779.

\bibitem{BFHS17}
H.~Babaee, M.~Farazmand, G.~Haller, T.~Sapsis, {R}educed-{O}rder {D}escription
  of {T}ransient {I}nstabilities and {C}omputation of {F}inite-time {L}yapunov
  {E}xponents, Chaos 27~(6) (2017) 063103.

\bibitem{SL09}
T.~Sapsis, P.~Lermusiaux, Dynamically orthogonal field equations for continuous
  stochastic dynamical systems, Physica D: Nonlinear Phenomena 238~(23-24)
  (2009) 2347--2360.

\bibitem{CHZI13}
M.~Cheng, T.~Hou, Z.~Zhang, {A} {D}ynamically {B}i-orthogonal {M}ethod for
  {T}ime-dependent {S}tochastic {P}artial {D}ifferential {E}quations {I}:
  {D}erivation and {A}lgorithms, J. Comput. Phys. 242 (2013) 843--868.

\bibitem{Babaee:2017aa}
H.~Babaee, M.~Choi, T.~Sapsis, G.~Karniadakis, {A} {R}obust
  {B}i-orthogonal/{D}ynamically-orthogonal {M}ethod using the {C}ovariance
  {P}seudo-inverse with {A}pplication to {S}tochastic {F}low {P}roblems, J.
  Comput. Phys. 344 (2017) 303--319.

\bibitem{B19}
H.~Babaee, {A}n {O}bservation-driven {T}ime-dependent {B}asis for a {R}educed
  {D}escription of {T}ransient {S}tochastic {S}ystems, Proc. R. Soc. Lond. A
  475~(2231) (2019) 20190506.

\bibitem{PB20}
P.~Patil, H.~Babaee, {R}eal-time {R}educed-order {M}odeling of {S}tochastic
  {P}artial {D}ifferential {E}quations via {T}ime-dependent {S}ubspaces, J.
  Comput. Phys. 415 (2020) 109511.

\bibitem{RNB21}
D.~Ramezanian, A.~Nouri, H.~Babaee, {O}n-the-fly {R}educed {O}rder {M}odeling
  of {P}assive and {R}eactive {S}pecies via {T}ime-{D}ependent {M}anifolds,
  arXiv preprint arXiv:2101.03847.

\bibitem{ZT03}
I.~Zs{\'e}ly, T.~Tur{\'a}nyi, {T}he {I}nfluence of {T}hermal {C}oupling and
  {D}iffusion on the {I}mportance of {R}eactions: {T}he {C}ase {S}tudy of
  {H}ydrogen--{A}ir {C}ombustion, Phys. Chem. Chem. Phys 5~(17) (2003)
  3622--3631.

\bibitem{ZZT03}
I.~Zsely, J.~Zador, T.~Turanyi, {S}imilarity of {S}ensitivity {F}unctions of
  {R}eaction {K}inetic {M}odels, J. Phys. Chem. A 107~(13) (2003) 2216--2238.

\bibitem{Williams1985}
F.~A. Williams, Turbulent combustion, in: J.~D. Buckmaster (Ed.), The
  Mathematics of Combustion, SIAM, Philadelphia, PA, 1985.

\bibitem{Cantera}
D.~Goodwin, H.~Moffat, R.~Speth, {C}antera: {A}n {O}bject-oriented {S}oftware
  {T}oolkit for {C}hemical {K}inetics, {T}hermodynamics, and {T}ransport
  {P}rocesses, \url{http://www.cantera.org}, {V}ersion 2.3.0 (2017).
\newblock \href {http://dx.doi.org/10.5281/zenodo.170284}
  {\path{doi:10.5281/zenodo.170284}}.

\bibitem{Burke12}
M.~Burke, M.~Chaos, Y.~Ju, F.~Dryer, S.~Klippenstein, {C}omprehensive {H}2/{O}2
  {K}inetic {M}odel for {H}igh-{P}ressure {C}ombustion, Int. J. Chem. Kinet.
  44~(7) (2012) 444--474.

\bibitem{SAM}
C.~Zhou, Y.~Li, E.~O'connor, K.~Somers, S.~Thion, C.~Keesee, O.~Mathieu,
  E.~Petersen, T.~DeVerter, M.~Oehlschlaeger, et~al., {A} {C}omprehensive
  {E}xperimental and {M}odeling {S}tudy of {I}sobutene {O}xidation, Combust.
  Flame 167 (2016) 353--379.

\bibitem{PPS18}
G.~Pio, V.~Palma, E.~Salzano, {C}omparison and {V}alidation of {D}etailed
  {K}inetic {M}odels for the {O}xidation of {L}ight {A}lkenes, Ind. Eng. Chem.
  Res. 57~(21) (2018) 7130--7135.

\bibitem{Luo12}
Z.~Luo, C.~Yoo, E.~Richardson, J.~Chen, C.~Law, T.~Lu, {C}hemical {E}xplosive
  {M}ode {A}nalysis for a {T}urbulent {L}ifted {E}thylene {J}et {F}lame in
  {H}ighly-{H}eated {C}oflow, Combust. Flame 159~(1) (2012) 265--274.

\end{thebibliography}

\end{document}